\documentclass[aps,prl,twocolumn,superscriptaddress,longbibliography]{revtex4-1}

\usepackage{amsmath}
\usepackage{amssymb}
\usepackage{graphicx}
\usepackage[colorlinks=true,
pdfborder={0 0 0},
linkcolor=blue,
citecolor=black]{hyperref}

\begin{document}
\title{Detecting chirality in two-terminal electronic devices}

\author{Xu~Yang}
\email[]{xu.yang@rug.nl}
\affiliation{Zernike Institute for Advanced Materials, University of Groningen, NL-9747AG Groningen, The Netherlands}

\author{Caspar~H.~van~der~Wal}
\affiliation{Zernike Institute for Advanced Materials, University of Groningen, NL-9747AG Groningen, The Netherlands}

\author{Bart~J.~van~Wees}
\affiliation{Zernike Institute for Advanced Materials, University of Groningen, NL-9747AG Groningen, The Netherlands}

%\date{\today}

\begin{abstract}
	Central to spintronics is the interconversion between electronic charge and spin currents, and this can arise from the chirality-induced spin selectivity (CISS) effect. CISS is often studied as magnetoresistance (MR) in two-terminal (2T) electronic devices containing a chiral (molecular) component and a ferromagnet. However, fundamental understanding of when and how this MR can occur is lacking. Here, we uncover an elementary mechanism that generates such a MR for nonlinear response. It requires energy-dependent transport and energy relaxation within the device. The sign of the MR depends on chirality, charge carrier type, and bias direction. Additionally, we reveal how CISS can be detected in the linear response regime in magnet-free 2T devices, either by forming a chirality-based spin-valve using two or more chiral components, or by Hanle spin precession in devices with a single chiral component. Our results provide operation principles and design guidelines for chirality-based spintronic devices and technologies.
\end{abstract}

\maketitle

Recognizing and separating chiral enantiomers using electronic/spintronic technologies addresses fundamental questions of electronic charge and spin transport~\cite{barron1986true_1}. It can open up new avenues for chiral chemistry, and can bring chiral (molecular) structures into electronic and spintronic applications. This is enabled by the chirality-induced spin selectivity (CISS) effect~\cite{naaman2019chiral,gohler2011spin}, which describes the generation of a collinear spin current by a charge current through a chiral component (single molecule, assembly of molecules, or solid-state system). In two-terminal (2T) electronic devices that contain a chiral component and a single ferromagnet (FM), CISS is reported as a change of (charge) resistance upon magnetization reversal~\cite{suda2019light,lu2019spin,aragones2017measuring,kiran2016helicenes,xie2011spin,torres2019room,liu2019spin,kulkarni2020highly}. This magnetoresistance (MR) has been interpreted in analogy to that of a conventional spin valve, based on the understanding that both the FM and the chiral component act as spin--charge converters~\cite{michaeli2015origin,medina2015continuum,guo2012spin,zollner2019chiral,diaz2018thermal,geyer2019chirality}. However, this interpretation overlooks the fundamental distinction between their underlying mechanisms -- magnetism breaks time reversal symmetry, while CISS, as a spin-orbit effect, does not. In fact, Onsager reciprocity prohibits the detection of spin-orbit effects as 2T charge signals using a single FM in the linear response regime~\cite{onsager1931reciprocal,buttiker1988symmetry,adagideli2006detection,yang2019spin}. Therefore, it requires theories beyond linear response for possible explanations of the MR observed in CISS experiments~\cite{yang2019spin,dalum2019theory,naaman2020comment,yang2020reply}.

Here we show that such a 2T MR can indeed arise from the breaking of Onsager reciprocity in the nonlinear regime. When the conditions for generating CISS in a chiral component are fulfilled, the emergence of the 2T MR requires two key ingredients: (a) energy-dependent electron transport due to, for instance, tunneling or thermally activated conduction through molecular orbitals; and (b) energy relaxation due to inelastic processes. Note that there exists an interesting parallel in chemistry, where absolute asymmetric synthesis is enabled by the lone influence of a magnetic field or magnetization in the nonlinear regime (details see Supplementary Information A [SI.A])~\cite{barron1986true_1,micali2012selection,banerjee2018separation}.

Below, before demonstrating the emergence of 2T MR in the nonlinear regime, we will first introduce a transport-matrix formalism unifying the description of coupled charge and spin transport in spin--charge converters such as a chiral component and a FM tunnel junction/interface (FMTJ). Afterwards, we will explore new device designs that make 2T electrical detection of CISS possible in the linear response regime, and reveal a chiral spin valve built without magnetic materials.

\section{A unified formalism for coupled charge and spin transport}
The (spin-resolved) Landauer formula considers charge voltages as the driving forces for electronic charge and spin transport~\cite{buttiker1985generalized}. However, in a circuit with multiple spin--charge converters, we must also consider the build-up of spin accumulations, which also drive the coupled charge and spin transport (details see SI.B)~\cite{brataas2000finite,shekhter2014mechanically}. We take this into account using a transport matrix formalism.

\subsection{Spin--charge conversion in a chiral component}
\begin{figure}[htp]
	\includegraphics[width=\linewidth]{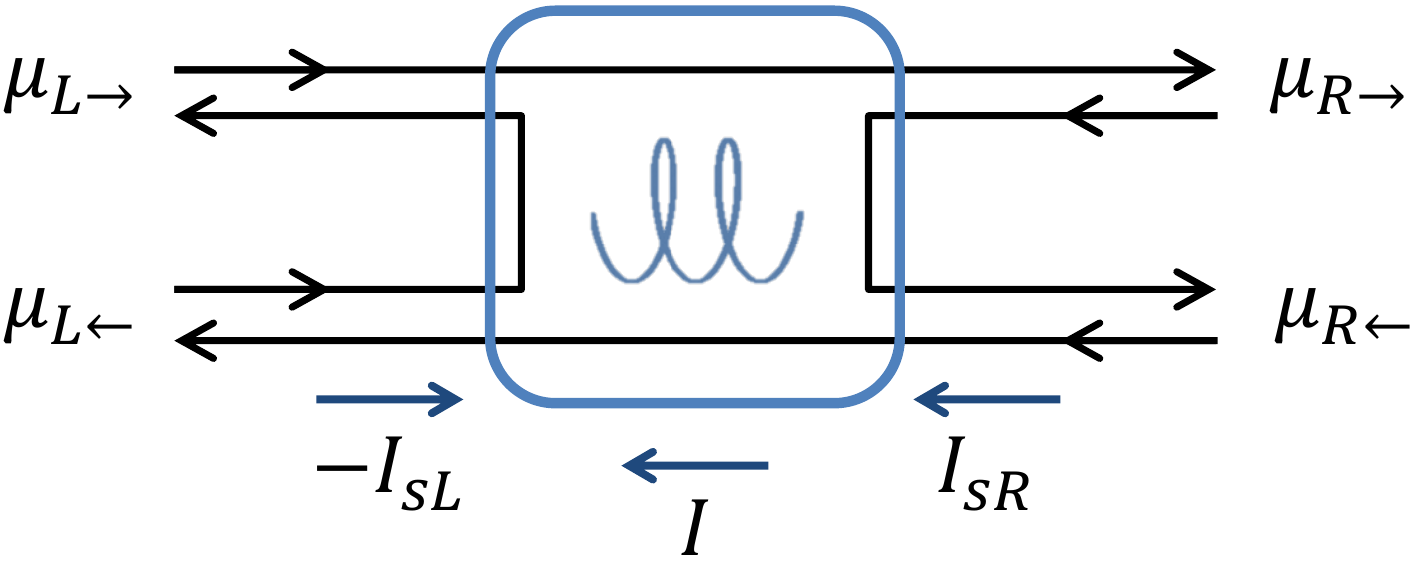}
	\caption{\label{fig:molecule}Illustration of CISS (ideal case). The directional electron transmission is spin-selective, and the unfavored spin is flipped and reflected. The chiral component is indicated by the blue helix, and is assumed to favor the transmission of electrons with spin parallel to momentum. The electrons on both sides ($L$ and $R$) of the chiral component are labeled with their spin-specific ($\rightarrow$ or $\leftarrow$) electrochemical potentials $\mu_{L(R)\rightarrow(\leftarrow)}$. At thermodynamic equilibrium, any net charge or spin current in and out of electrodes is forbidden, but when biased, the chiral structure supports a charge current $I$ and collinear spin currents on both sides $I_{sL}$ and $I_{sR}$. The positive currents are defined as right-to-left, i.e. when the (spin-polarized) electrons flow from left to right.}
\end{figure}

CISS arises from spin-orbit interaction and the absence of space-inversion symmetry. Further symmetry considerations require that the sign of the CISS-induced collinear spin currents must depend on the direction of the charge current and the (sign of) chirality. Note that the generation of CISS requires a nonunitary transport mechanism within the chiral component~\cite{bardarson2008proof,matityahu2016spin}, which we assume to be present (details see SI.C). In an ideal case, as illustrated in Fig.~\ref{fig:molecule}, this directional, spin-dependent electron transport effectively allows only one spin orientation, say, parallel to the electron momentum, to transmit through the chiral component, and reflects and spin-flips the other~\cite{yang2019spin}. The spin-flip reflection prevents a net spin current in and out of electrodes at thermodynamic equilibrium~\cite{kiselev2005prohibition}.

We first describe the directional electron transmission ($\mathbb{T}$) and reflection ($\mathbb{R}$) in a generalized chiral component using spin-space matrices introduced in Ref.~\onlinecite{yang2019spin}. For right-moving (subscript $\rhd$) electrons coming from the left-hand side of the component
\begin{equation}\label{eqn:matrixmolR}
\mathbb{T}_{\rhd}=\begin{pmatrix}
t_{\rightarrow\rightarrow} & t_{\leftarrow\rightarrow} \\
t_{\rightarrow\leftarrow} & t_{\leftarrow\leftarrow}
\end{pmatrix},\; \;
\mathbb{R}_{\rhd}=\begin{pmatrix}
r_{\rightarrow\rightarrow} & r_{\leftarrow\rightarrow}\\
r_{\rightarrow\leftarrow} & r_{\leftarrow\leftarrow}
\end{pmatrix},
\end{equation}
where the matrix elements are probabilities of an electron being transmitted ($t$) or reflected ($r$) from an initial spin state (first subscript) to a final spin state (second subscript). For left-moving electrons, the corresponding matrices are the time-reversed forms of the above (details see SI.D).

We extend the above coupled charge and spin transport by converting the spin-space matrices to transport matrices that link the thermodynamic drives and responses in terms of charge and spin. We define (charge) electrochemical potential $\mu=(\mu_{\rightarrow}+\mu_{\leftarrow})/2$ and spin accumulation $\mu_s=(\mu_{\rightarrow}-\mu_{\leftarrow})/2$, as well as charge current $I=I_{\rightarrow}+I_{\leftarrow}$ and spin current $I_s=I_{\rightarrow}-I_{\leftarrow}$. A subscript $R$ or $L$ is added when describing the quantities on a specific side of the component. With these, for a generalized case of Fig.~\ref{fig:molecule}, we derive (details see SI.D)
\begin{equation}\label{eqn:molmatfull}
\begin{pmatrix}
I\\-I_{sL}\\I_{sR}
\end{pmatrix}=-\frac{Ne}{h}\begin{pmatrix}
t & s & s \\
P_r r & \gamma_r & \gamma_t \\
P_t t & \gamma_t & \gamma_r
\end{pmatrix} \begin{pmatrix}
\mu_L-\mu_R \\ \mu_{sL} \\ \mu_{sR}
\end{pmatrix},
\end{equation}
where $N$ is the number of (spin-degenerate) channels, $e$ is elemental charge (positive value), and $h$ is the Planck's constant. We name the $3\times3$ matrix the charge--spin transport matrix $\mathcal{T}$. All its elements are linear combinations of the spin-space $\mathbb{T}$ and $\mathbb{R}$ matrix elements, and represent key transport properties of the chiral component. For example, $t$ is the (averaged) transmission probability, $r$ is the reflection probability (note that $t+r=2$ because we have treated the two spins separately), $P_{t}$ and $P_{r}$ are the CISS-induced spin polarizations of the transmitted and reflected electrons, respectively. An electrochemical potential difference $\mu_L-\mu_R=-eV$ is provided by an bias voltage $V$. 

This matrix equation fully describes the coupled charge and collinear spin transport through a (nonmagnetic) chiral component, which is subject to Onsager reciprocity in the linear response regime. This requires $\mathcal{T}_{ij}(\boldsymbol{H},\boldsymbol{M})=\mathcal{T}_{ji}(-\boldsymbol{H},-\boldsymbol{M})$, where $\boldsymbol{H}$ is the magnetic field and $\boldsymbol{M}$ is the magnetization~\cite{onsager1931reciprocal}. This then gives $P_t t=P_r r=s$. In later discussions we will connect the $R$-side of the chiral component to an electrode (reservoir), where $\mu_{sR}=0$ and $I_{sR}$ is irrelevant. The $\mathcal{T}$ matrix then reduces to a $2\times2$ form (details see SI.D)
\begin{equation}\label{eqn:molmatmin}
\begin{pmatrix}
I\\-I_{sL}
\end{pmatrix}=-\frac{Ne}{h}\begin{pmatrix}
t & P_t t \\
P_t t & \gamma_r
\end{pmatrix} \begin{pmatrix}
\mu_L-\mu_R \\ \mu_{sL} 
\end{pmatrix}.
\end{equation}
Note that $t$ and $\gamma_r$ do not depend on the (sign of) chirality, while $P_t$ changes sign when the chirality is reversed.

\subsection{Spin--charge conversion in a FMTJ}
To calculate the coupled charge and spin transport in a generic 2T circuit where an (achiral) ferromagnet is also present, we need to derive a similar $\mathcal{T}$ matrix for a FMTJ. A FM breaks time-reversal symmetry and provides a spin-polarization $P_{FM}$ to any outflowing charge current. Based on this, we obtain for the $R$-side of the FMTJ (details see SI.E)
\begin{equation}\label{eqn:fmmatlin}
\begin{pmatrix}
I\\I_{sR}
\end{pmatrix}=-\frac{N'e}{h}\begin{pmatrix}
T & -P_{FM} T \\
P_{FM} T & -T
\end{pmatrix} \begin{pmatrix}
\mu_L -\mu_R \\ \mu_{sR}
\end{pmatrix},
\end{equation}
where $N'$ is the number of (spin-degenerate) channels, and $T$ is the electron transmission probability accounting for both spins.

The matrix $\mathcal{T}$ here also satisfies $\mathcal{T}_{ij}(\boldsymbol{H},\boldsymbol{M})=\mathcal{T}_{ji}(-\boldsymbol{H},-\boldsymbol{M})$, where a reversal of $\boldsymbol{M}$ corresponds to a sign change of $P_{FM}$.

\section{Origin of MR -- energy-dependent transport and energy relaxation}
\subsection{No MR in the linear response regime}
\begin{figure}[htp]
	\includegraphics[width=\linewidth]{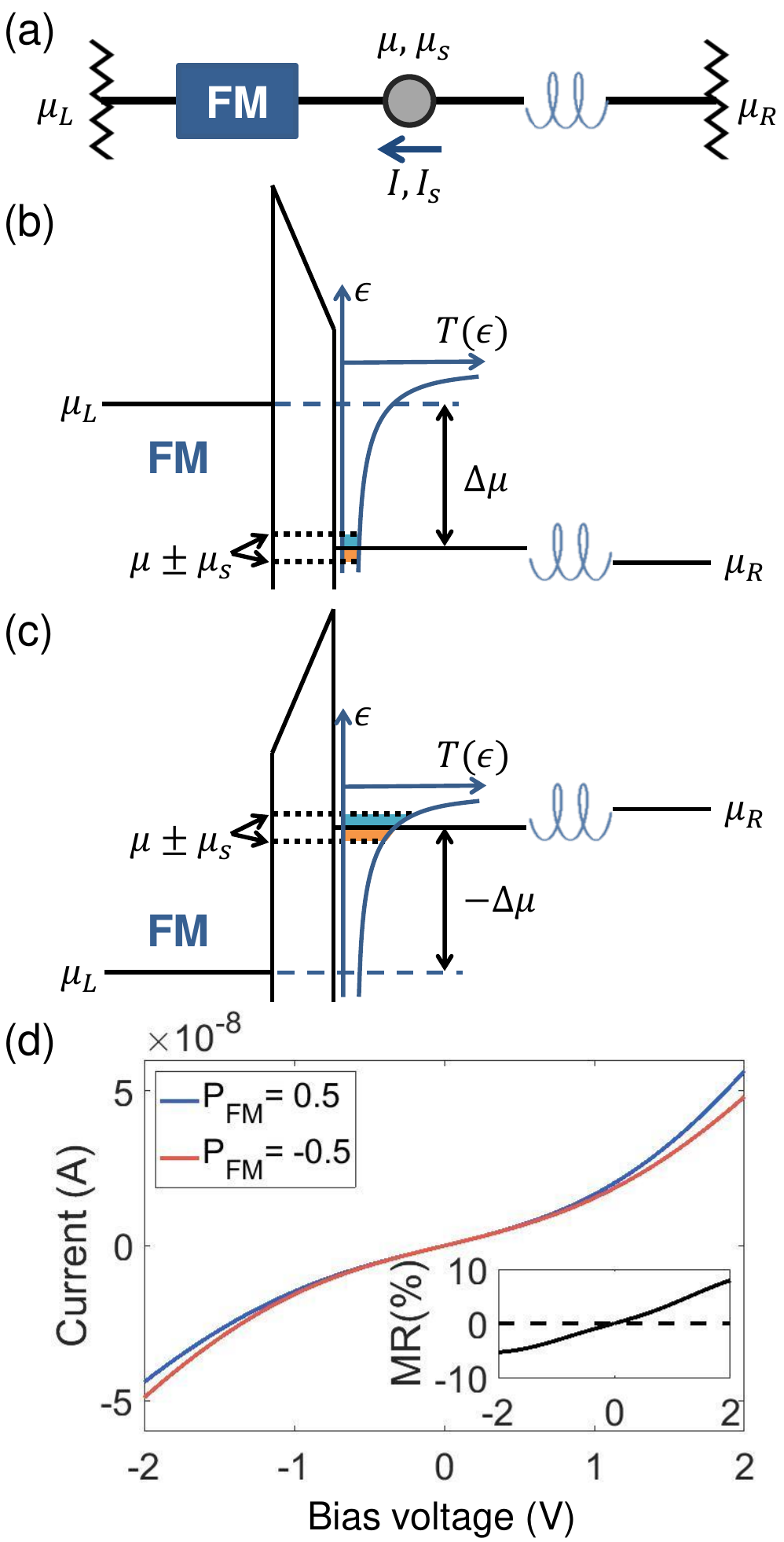}
	\caption{\label{fig:circuit}Origin of MR in a generic 2T circuit. (a). A 2T circuit containing a FM and a chiral component, connected by a node. The chiral component is assumed to favor the transmission of electrons with spin parallel to momentum. (b),(c). Schematic energy diagrams of tunneling at the FMTJ under forward [(b)] and reverse [(c)] biases. The energy-dependent tunnel transmission $T(\epsilon)$ is sketched in blue, and the tunnel current affected by the spin accumulation $\mu_s$ is illustrated by the color-shaded areas (cyan and orange). (d). Example $I$-$V$ curves considering the nonlinear mechanism of (b)-(c), while keeping the transmission through the chiral component constant (details see SI.G). Note that positive bias voltage corresponds to the reverse bias scenario depicted in panel (c). The MR ratio, defined as $(I_+-I_-)/(I_++I_-)$, is plotted in inset. Here the subscript $+$ or $-$ denotes the corresponding sign of $P_{FM}$, with $P_{FM}>0$ corresponding to a spin-right polarization for the injected electrons. The dashed line marks zero MR.} 
\end{figure}

We model a generic 2T MR measurement geometry using Fig.~\ref{fig:circuit}(a). A FM and a chiral component are connected in series between two spin-unpolarized electrodes ($L$ and $R$), and the difference between their electrochemical potentials $\mu_L-\mu_R=-eV$ drives charge and spin transport. We introduce a node between the FM and the chiral component, which is characterized by an electrochemical potential $\mu$ and a spin accumulation $\mu_s$. It preserves the spin but relaxes the energy of electrons to a Fermi-Dirac distribution due to inelastic processes (e.g. electron-phonon interaction).

By adapting Eqn.~\ref{eqn:molmatmin} and Eqn.~\ref{eqn:fmmatlin} in accordance with the notations in Fig.~\ref{fig:circuit}(a), and applying continuity condition in the node for both the charge and spin currents, we derive the 2T conductance for the linear response regime (details see SI.F)
\begin{equation}\label{eqn:g2t}
G_{2T}=G_{2T} \left(P^2_{FM}, P^2_{t} \right), \text{~no~} P_{FM}P_t \text{~term.} 
\end{equation}
It depends on the polarizations $P_{FM}$ and $P_t$ only to second order, and does not depend on their product $P_{FM}P_t$. Therefore, the 2T conductance remains unchanged when the sign of either $P_{FM}$ or $P_t$ is reversed by the reversal of either the FM magnetization direction or the chirality. This result again confirms the vanishing MR in the linear response regime, as strictly required by Onsager reciprocity~\cite{yang2019spin}.  

This vanishing MR can be understood as a result of two simultaneous processes. First, by the conventional description, the charge current through the chiral component drives a collinear spin current and creates a spin accumulation in the node (spin injection by CISS), which is then detected as a charge voltage by the FM (spin detection by FM). This charge voltage indeed changes upon the FM magnetization reversal. However, this is always accompanied by the second process, where it is the FM that injects a spin current, and the Onsager reciprocal of CISS detects it as a charge voltage. This voltage also changes upon the FM magnetization reversal. In the linear regime, the two processes compensate each other, and the net result is a zero MR.

\subsection{Emergence of MR in nonlinear regime}
The key to inducing MR is to break the balance between the two processes, which can be done by using electrons at different energies for spin injection and spin detection. This requires the presence of energy relaxation inside the device, and it also needs the transport to be energy-dependent in at least one of the spin--charge converters. Note that the energy relaxation is crucial for generating MR, because Onsager reciprocity, which holds at each energy level, would otherwise prevent a MR even in the nonlinear regime despite the energy-dependent transport. We illustrate the emergence of MR using two types of energy dependence, (a) quantum tunneling through the FMTJ, and (b) thermally activated conduction through molecular orbitals.  

The asymmetric spin injection and spin detection in a FMTJ was previously discussed by Jansen \textit{et al.} for a FM coupled to a semiconductor~\cite{jansen2018nonlinear}, and here we generalize it for our system. The energy diagrams for this tunneling process are sketched in Fig.~\ref{fig:circuit}(b)-(c) for opposite biases. The bias opens up an energy window $\Delta \mu=\mu_L-\mu$, and the electrons within this energy window contribute to the total charge current $I$. The energy distribution of the tunneling electrons follows the energy dependence of the tunnel transmission probability $T(\epsilon)$ (blue curve). Therefore, the electrons that contribute the most to the tunnel current $I$ are those at the highest available energy, which are at $\mu_L$ (in the FM) under forward bias, and are at the spin-split electrochemical potentials $\mu \pm \mu_s$ (in the node) under reverse bias. 

The spin injection process concerns the spin current $I_s$ induced by the total charge current $I$ through the FMTJ with spin polarization $P_{FM}$ (assuming energy-independent $P_{FM}$). It is determined by the energy integral of $T(\epsilon)$ over the entire bias-induced window $\Delta \mu$, and is symmetric for opposite biases (assuming $\mu_s \ll \Delta\mu$). In contrast, the spin detection process, which concerns the spin accumulation $\mu_s$ in the node at its highest energy $\mu$  (Fermi level), is not symmetric for opposite biases. Effectively, the spin accumulation describes the deficit of one spin and the surplus of the other, as illustrated by the orange- and blue-shaded regions under the $T(\epsilon)$ curve in Fig.~\ref{fig:circuit}(b)-(c), and therefore drives an (additional) charge current proportional to the area difference between the two regions. This detected charge current depends on the transmission probability at energy $\mu$, and increases monotonically as the bias becomes more reverse.

\begin{figure}[htp]
	\includegraphics[width=\linewidth]{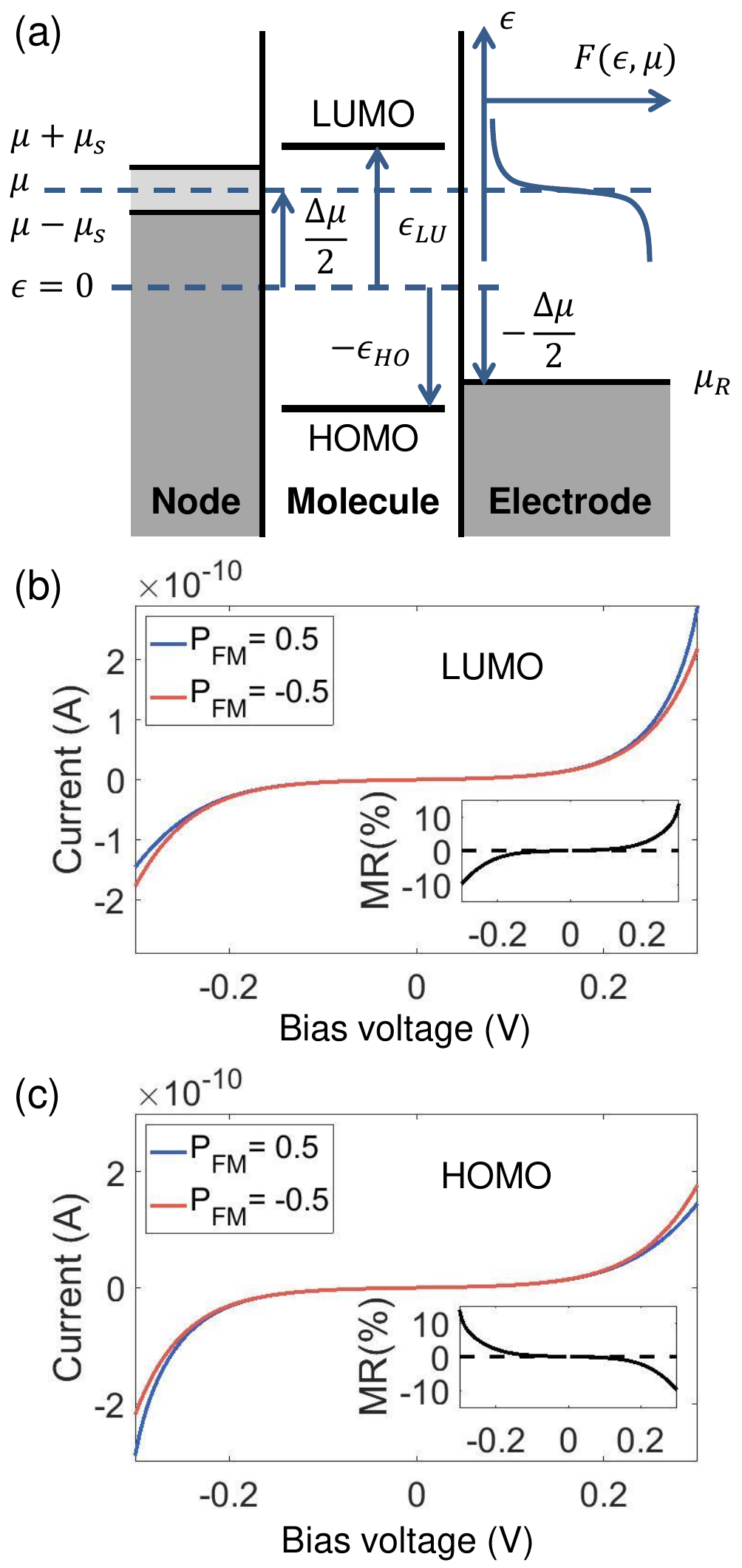}
	\caption{\label{fig:circuit_iv}Generating MR by thermally activated conduction through molecular orbitals. (a). Schematic energy diagram of resonant transmission through molecular orbitals in a chiral component (molecule). The LUMO and HOMO levels and the bias-dependent electrochemical potentials are labeled, and the energy- and bias-dependent Fermi-Dirac function $F(\epsilon,\mu)$ is sketched in blue. (b)-(c). Example $I$-$V$ curves and MR (inset) due to the resonant transmission through the LUMO [(b)] and the HOMO [(c)], for the same device geometry as in Fig.~\ref{fig:circuit}(a) but with the transmission of the FMTJ set constant. The chiral molecule is assumed to favor the transmission of electrons with spin parallel to momentum.}
\end{figure}

The different bias dependences for spin injection and detection break Onsager reciprocity for nonlinear response. This is also shown by the different off-diagonal terms in the nonlinear transport equation (details see SI.G)
\begin{equation}\label{eqn:fmmatnonlin}
\begin{pmatrix}
I\\ \hspace{-0.3cm} \\I_s
\end{pmatrix}=-\frac{N'e}{h}\begin{pmatrix}
\overline{T}|_{\mu}^{\mu_L} & -P_{FM}T|_{\epsilon=\mu} \\ \hspace{-0.3cm} \\
P_{FM}\overline{T}|_{\mu}^{\mu_L} & -T|_{\epsilon=\mu}
\end{pmatrix} \begin{pmatrix}
\mu_L - \mu \\ \hspace{-0.3cm} \\ \mu_s
\end{pmatrix},
\end{equation}
where $\overline{T}|_{\mu}^{\mu_L}=[1/(\mu_L-\mu)]\int_{\mu}^{\mu_L}T(\epsilon) d\epsilon$ is the averaged transmission over the energy window $\Delta \mu = \mu_L-\mu$, and $T|_{\epsilon=\mu}$ is the transmission evaluated at the Fermi level of the node $\epsilon=\mu$. In the linear response regime, when $\mu_L \approx \mu$, this equation returns to Eqn.~\ref{eqn:fmmatlin}. 

The tunnel $I$-$V$ and the MR due to this mechanism are illustrated in Fig.~\ref{fig:circuit}(d) using realistic circuit parameters (details see SI.I). The MR ratio reaches nearly $10~\%$ at large biases, but strictly vanishes at zero bias. Notably, the MR is positive under positive bias voltage (corresponds to reverse bias as in Fig.~\ref{fig:circuit}(c)), and it reverses sign as the bias changes sign.

The non-reciprocal spin injection and detection can also arise from the nonlinear transport through the chiral component. In principle, this could also be due to tunneling, but we focus here on another aspect, the Fermi-Dirac distribution of electrons. This is negligible when the transmission function $T(\epsilon)$ is smooth, as for the case of tunneling (thus we have assumed zero temperature for deriving Eqn.~\ref{eqn:fmmatnonlin}), but it becomes dominant when electron (or hole) transmission is only allowed at certain discrete energy levels or energy bands that are away from Fermi level, as for the case of conduction through molecular orbitals or through energy bands in semiconductors. We illustrate this in Fig.~\ref{fig:circuit_iv}(a) considering the resonant transmission through the LUMO (lowest unoccupied molecular orbital) and the HOMO (highest occupied molecular orbital) of a chiral molecule (details see SI.H). For spin injection, the generated spin current is proportional to the total charge current, which depends on the (bias-induced) electrochemical potential difference between the node and the right electrode, and is symmetric for opposite biases. In comparison, for spin detection, the spin-split electrochemical potentials in the node $\mu \pm \mu_s$ induce unequal occupations of opposite spins at each MO (depending on the MO position with respect to the node Fermi level $\mu$), and it is not symmetric for opposite biases. This different bias dependence breaks Onsager reciprocity for nonlinear response, and gives rise to MR. 

We consider the transmission through either only the LUMO or only the HOMO, and their example $I$-$V$ curves and MR ratios are plotted in Fig.~\ref{fig:circuit_iv}(b)-(c), respectively (details see SI.I). The MR is able to reach tens of percent even at relatively small biases, and changes sign as the bias reverses. Remarkably, the bias dependence of the MR is opposite for LUMO and HOMO, implying that the charge carrier type, i.e. electrons or holes, co-determines the sign of the MR. Again, the MR strictly vanishes as the bias returns to zero (linear response regime). An overview of the signs of MR is given in SI.J. 

\section{Chiral spin valve}
\begin{figure*}[htp]
	\includegraphics[width=0.7\linewidth]{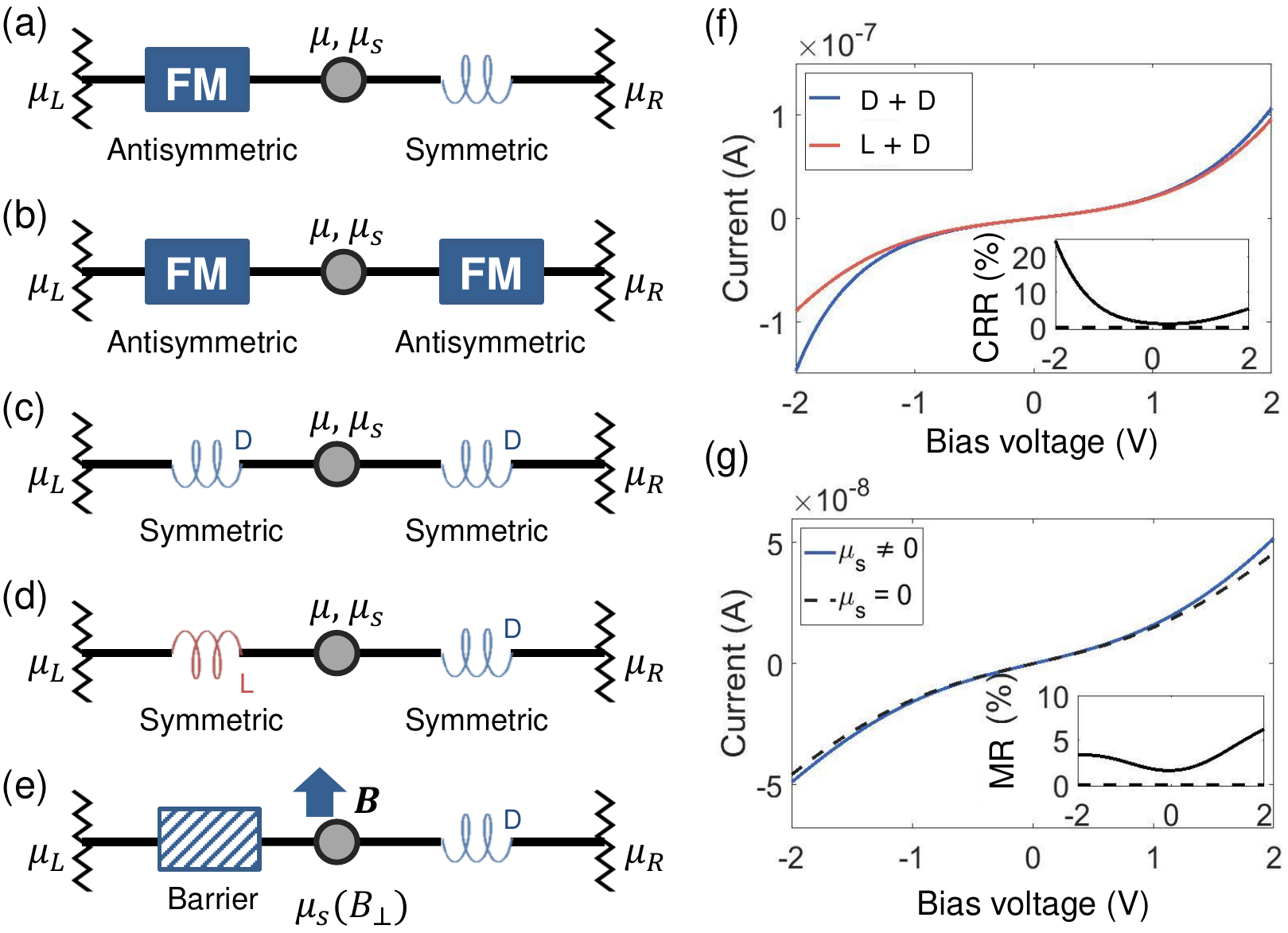}
	\caption{\label{fig:chiralsv}Generic 2T spin-valve device geometries with the symmetry of the charge--spin transport matrix labeled for each component. (a). The aforementioned FM--chiral geometry where MR signals are strictly forbidden in the linear response regime. (b). A FM--FM geometry, as in a conventional spin valve, where MR signals are allowed in the linear regime. (c),(d). A chiral--chiral geometry for using the same [(c)] and opposite [(d)] chiralities, as marked by color and labeled with \textsc{d} or \textsc{l} (here we assume the \textsc{d}-chiral component favors the transmission of electrons with spin parallel to momentum, and the \textsc{l}-chiral component favors the anti-parallel ones). The spin-valve effect can be achieved, even in the linear regime, by reversing the chirality of one component. (e). A geometry for directly probing the spin accumulation generated by a single chiral component. The perpendicular magnetic field $B$ suppresses spin accumulation in the node via Hanle spin precession. (f). Example $I$-$V$ curves for a chiral---chiral spin valve, with the two curves representing the geometries in panel (c) and (d) respectively. The corresponding chirality-reversal resistance (CRR) ratio, as defined by $CRR=(I_{\textsc{dd}}-I_{\textsc{ld}})/(I_{\textsc{dd}}+I_{\textsc{ld}})$ (the two subscripts refer to the chiralities of the two chiral components), is plotted in the inset. (g). Example $I$-$V$ for the geometry in panel (e), calculated for cases with $\mu_s$ either fully or not-at-all suppressed by Hanle precession. The corresponding MR is shown in inset, which is defined as the difference of the two curves divided by their sum.}
\end{figure*}

In the linear response regime, the coupled charge and spin transport at the FMTJ is described by an antisymmetric matrix (opposite off-diagonal terms), and for a chiral component the matrix is symmetric. The symmetries of these matrices are directly required by Onsager reciprocity, and have consequences when considering 2T circuits containing two generic spin--charge converting components, as illustrated in Fig.~\ref{fig:chiralsv}. As long as one of the two components is ferromagnetic (antisymmetric matrix) and the other is not (symmetric matrix), a 2T MR signal is forbidden in the linear response regime (Fig.~\ref{fig:chiralsv}(a)). The restriction is however not in place if both matrices have the same symmetry. An example is the well-known conventional spin valve, as shown in Fig.~\ref{fig:chiralsv}(b), where two FMs are connected in series. Both FMs are described by antisymmetric matrices, and the magnetization reversal of one FM indeed changes the 2T conductance in the linear response regime.

A less obvious outcome of this symmetry consideration is that a combination of two chiral components (both described by symmetric matrices) can also form a spin valve, provided that at least one of them can switch chirality, such as a molecular rotor~\cite{koumura1999light}. This is illustrated in Fig.~\ref{fig:chiralsv}(c)-(d), with example $I$-$V$ curves in Fig.~\ref{fig:chiralsv}(f). In the linear response regime, this geometry already produces a nonzero chirality-reversal resistance (CRR, see figure caption for definition) ratio, which is further enhanced to tens of percent as the bias increases.

Finally, we introduce an experimental geometry that can probe the spin injection caused by a single chiral component. As shown in Fig.~\ref{fig:chiralsv}(e), the chiral component is connected to a nonmagnetic tunnel barrier via a node. Through CISS, a charge current creates a spin accumulation in the node (even in the linear response regime), which can then be suppressed using a perpendicular magnetic field due to Hanle spin precession. This results in a magnetic-field-dependent 2T conductance, and Fig.~\ref{fig:chiralsv}(g) shows the $I$-$V$ curves for zero magnetic field (blue solid curve) and for when the field fully suppresses the spin accumulation (black dashed curve). The corresponding MR (inset) for this geometry is nonzero in the linear response regime, and can be enhanced by increasing bias. 

\section{Discussion}
We have explained that energy-dependent charge transport combined with energy relaxation provide a fundamental mechanism that breaks the Onsager reciprocity in the nonlinear regime, and therefore generates MR in a 2T circuit containing a FM and a chiral component. We separately discussed two sources of energy dependence, tunneling at the FMTJ and the thermally activated resonant transmission through molecular orbitals in a chiral component. In experimental conditions, the two effects can be simultaneously present. Also, the exact bias dependence may vary depending on the device details, for example, due to the presence of Schottky barriers or due to conduction through valence and conduction bands. 

We have assumed that the conduction mechanisms in each separate circuit component are not changed when building them together into one device, so that the overall conduction is (phase) incoherent. In practice, coherent conduction mechanisms, such as direct tunneling from the FM into a MO of the chiral molecule, may also be present. For fully incoherent conduction, the chirality-based spin valve (Fig.~\ref{fig:chiralsv}(c-d)) resembles a conventional CPP GMR device~\cite{baibich1988giant}, while if coherent tunneling dominates, it is comparable to a TMR device~\cite{moodera1995large}. 

Finally, we draw attention to the sign of the nonlinear MR, which depends on the dominating nonlinear element, the bias direction, the charge carrier type, and the (sign of) chirality, as summarized in SI.J. We have plotted the MR as a function of bias while assuming the chirality and the charge carrier type remain unchanged. However, in experimental conditions, it is possible that the charge carrier type switches when the bias is reversed, and consequently the MR can have the same sign for opposite biases. Such a bias-even MR is indeed in agreement with most experimental observations~\cite{suda2019light,lu2019spin,kiran2016helicenes,xie2011spin,kulkarni2020highly,torres2019room,liu2019spin}. The observations concerning the vanishing MR in the linear regime are however unclear, since for experiments where a linear regime can be identified, some do show a strongly suppressed or vanishing MR~\cite{kiran2016helicenes,kulkarni2020highly}, while some do not~\cite{torres2019room,liu2019spin}. This thus requires further investigation.

\section{Acknowledgments}
This work is supported by the Zernike Institute for Advanced Materials (ZIAM), and the Spinoza prize awarded to Professor B.~J.~van~Wees by the Nederlandse Organisatie voor Wetenschappelijk Onderzoek (NWO). The authors acknowledge discussions with R.~Naaman, C.~Herrmann, A.~Aharony, and G.E.W.~Bauer.

\end{document}

\title{Supplementary Information: Detecting chirality in two-terminal electronic devices}

\author{Xu~Yang}
\email[]{xu.yang@rug.nl}
\affiliation{Zernike Institute for Advanced Materials, University of Groningen, NL-9747AG Groningen, The Netherlands}

\author{Caspar~H.~van~der~Wal}
\affiliation{Zernike Institute for Advanced Materials, University of Groningen, NL-9747AG Groningen, The Netherlands}

\author{Bart~J.~van~Wees}
\affiliation{Zernike Institute for Advanced Materials, University of Groningen, NL-9747AG Groningen, The Netherlands}

%\date{\today}

\maketitle

\vfill
\widetext

\tableofcontents
\pagebreak

\setcounter{equation}{0}
\setcounter{figure}{0}
\setcounter{page}{1}
\setcounter{section}{0}
\makeatletter
\renewcommand{\theequation}{S\arabic{equation}}
\renewcommand{\thefigure}{S\arabic{figure}}
\renewcommand{\thetable}{S\Roman{table}}
\renewcommand{\bibnumfmt}[1]{[S#1]}
\renewcommand{\citenumfont}[1]{S#1}

\newpage

\section{A. Comparison to absolute asymmetric (chemical) synthesis}
Absolute asymmetric synthesis refers to chemical reactions starting from achiral or racemic reactants but yield chiral products with a net enantiomeric excess. Generally, it requires the presence of a truly chiral external influence, which, according to Barron~\cite{barron1986true_1}, is a physical quantity that can appear in two degenerate forms (e.g. opposite signs of one physical quantity, or parallel and anti-parallel alignments of two vectors), which are interconverted by space-inversion but not time-reversal. Following this, the lone influence of a magnetic field, which reverses sign under time-reversal but not space-inversion, is not truly chiral, and therefore should not induce absolute asymmetric synthesis. 

Just like Onsager reciprocity, this true-chirality consideration is based on microscopic reversibility, and therefore is only strict in the linear response regime (in the vicinity of thermodynamic equilibrium). Barron later pointed out that~\cite{barron1987reactions}, in a chemical reaction that would yield a racemic mixture (of enantiomers), the presence of a magnetic field allows different reaction rates toward opposite enantiomers. This creates an enantiomeric excess away from equilibrium, which should eventually disappear as the system returns to equilibrium. However, if the nonequilibrium enantiomeric excess is amplified, for example by self-assembly or crystallization, absolute asymmetric synthesis is induced, and this was later demonstrated experimentally~\cite{micali2012selection,banerjee2018separation}. 

This shows that a time-reversal-breaking influence (the magnetic field), albeit not being truly chiral, can distinguish and separate enantiomers away from thermodynamic equilibrium. By the same token, in the 2T circuit we discuss, the FM breaks time reversal symmetry and can therefore indeed distinguish enantiomers away from equilibrium, i.e. in the nonlinear regime. Therefore, the nonlinear conductance of the 2T circuit may depend on the chirality of the circuit component, as well as the magnetization direction of the FM.

\section{B. Beyond the Landauer formula}
The Landauer formula describes the electrical conductance of a conductor using its scattering properties~\cite{landauer1989conductance}. Particularly, it expresses conductance in terms of transmission probabilities between electrodes, but does not require the use of reflection probabilities. Moreover, it considers that the charge (and spin) transport are solely driven by a charge bias, and neglects possible spin accumulations in the circuit~\cite{buttiker1985generalized}. For a 2T device, it allows two independent parameters (charge and spin transmission probability) for the description of coupled charge and spin transport.

We point out here that, a full description of the coupled charge and spin transport must extend beyond the conventional spin-resolved Landauer formula, and include also the (spin-flip) reflection terms and the build-up of spin accumulations, which also act as driving forces of the transport. This point was also earlier raised for a FM--normal metal system~\cite{brataas2000finite}. Following this, we need to characterize the (twofold rotationally symmetric) chiral component using four independent parameters, see the matrix in Eqn.~\ref{eqn:molmatwide}, contrasting to the two parameters mentioned earlier.

\section{C. Nonunitarity for generating CISS and energy relaxation for generating MR}
The generation of CISS requires the presence of nonunitary effects inside the chiral component, because the Kramers degeneracy would otherwise require equal transmission probabilities for opposite spin orientations~\cite{bardarson2008proof,matityahu2016spin}. These nonunitary effects break the phase information of a transmitting electron, but does not necessarily alter its energy.

The presence of these nonunitary effects only allows the generation of CISS, but does not guarantee the generation of a 2T MR. These nonunitary effects do not (necessarily) break the Onsager reciprocity, and therefore a 2T MR is still forbidden in the linear response regime. Even when considering energy-dependent transport in the nonlinear regime ($|eV|>k_B T$), since at each energy level the Onsager reciprocity still holds, a 2T MR cannot arise.

The emergence of MR, as we discussed, requires energy relaxation in the device. These relaxation processes not only break the phase information, but also alter the energy of the electrons, and rearrange them according to Fermi-Dirac distribution. We have assumed that they do not change the spin orientation of the electrons. In principle, we can also include spin relaxation in the node, which we expect to reduce the MR without changing its sign.  

\section{D. Spin and charge transport in a nonmagnetic chiral component}
In the main text we introduced the spin-space transmission and reflection matrices for the right-moving electrons in a (nonmagnetic) chiral component (Eqn.~1)
\begin{equation}
\mathbb{T}_{\rhd}=\begin{pmatrix}
t_{\rightarrow\rightarrow} & t_{\leftarrow\rightarrow} \\
t_{\rightarrow\leftarrow} & t_{\leftarrow\leftarrow}
\end{pmatrix},\; \;
\mathbb{R}_{\rhd}=\begin{pmatrix}
r_{\rightarrow\rightarrow} & r_{\leftarrow\rightarrow}\\
r_{\rightarrow\leftarrow} & r_{\leftarrow\leftarrow}
\end{pmatrix}.
\end{equation}

For the left-moving electrons, the matrices are the time-reversed form of the above
\begin{equation}\label{eqn:matrixmolL}
\mathbb{T}_{\lhd}=\begin{pmatrix}
t_{\leftarrow\leftarrow} & t_{\rightarrow\leftarrow} \\
t_{\leftarrow\rightarrow} & t_{\rightarrow\rightarrow}
\end{pmatrix},\; \;
\mathbb{R}_{\lhd}=\begin{pmatrix}
r_{\leftarrow\leftarrow} & r_{\rightarrow\leftarrow}\\
r_{\leftarrow\rightarrow} & r_{\rightarrow\rightarrow}
\end{pmatrix}.
\end{equation}
Note that these matrices are not suitable for describing magnetic components where time-reversal symmetry is not preserved, and we have assumed the chiral component is symmetric (i.e. a twofold rotational symmetry with axis perpendicular to the electron pathway, this ensures that for oppositely moving electrons, the spin polarization only changes sign).

We use spin-space column vector to describe electrochemical potentials and currents on both sides of the molecule, and following Ref.~\onlinecite{yang2019spin} we have
\begin{subequations}\label{eqn:currspins}
	\begin{align}
	\begin{pmatrix}
	I_{L\rightarrow}\\I_{L\leftarrow}
	\end{pmatrix}&=-\frac{Ne}{h} \left[ \left(\mathbb{I}-\mathbb{R}_{\rhd} \right) \begin{pmatrix}
	\mu_{L\rightarrow}\\\mu_{L\leftarrow}
	\end{pmatrix}  - \mathbb{T}_{\lhd}  \begin{pmatrix}
	\mu_{R\rightarrow}\\\mu_{R\leftarrow}
	\end{pmatrix} \right], \\
	-\begin{pmatrix}
	I_{R\rightarrow}\\I_{R\leftarrow}
	\end{pmatrix}&=-\frac{Ne}{h} \left[ \left(\mathbb{I}-\mathbb{R}_{\lhd} \right) \begin{pmatrix}
	\mu_{R\rightarrow}\\\mu_{R\leftarrow}
	\end{pmatrix}  - \mathbb{T}_{\rhd}  \begin{pmatrix}
	\mu_{L\rightarrow}\\\mu_{L\leftarrow}
	\end{pmatrix} \right],
	\end{align}
\end{subequations}
where $N$ is the number of spin-degenerate channels.

We define charge electrochemical potential $\mu=(\mu_{\rightarrow}+\mu_{\leftarrow})/2$ and spin accumulation $\mu_s=(\mu_{\rightarrow}-\mu_{\leftarrow})/2$, as well as charge current $I=I_{\rightarrow}+I_{\leftarrow}$ and spin current $I_s=I_{\rightarrow}-I_{\leftarrow}$. We will describe both charge and spin in electrical units.

Following these definitions, we have
\begin{subequations}\label{eqn:convspinpot}
	\begin{align}
	I&=I_{L\rightarrow}+I_{L\leftarrow}=I_{R\rightarrow}+I_{R\leftarrow},\\
	I_{sL}&=I_{L\rightarrow}-I_{L\leftarrow},\\
	I_{sR}&=I_{R\rightarrow}-I_{R\leftarrow},\\
	\mu_L&=(\mu_{L\rightarrow}+\mu_{L\leftarrow})/2,\\
	\mu_R&=(\mu_{R\rightarrow}+\mu_{R\leftarrow})/2,\\
	\mu_{sL}&=(\mu_{L\rightarrow}-\mu_{L\leftarrow})/2,\\
	\mu_{sR}&=(\mu_{R\rightarrow}-\mu_{R\leftarrow})/2.
	\end{align}
\end{subequations}
Combining Eqn.~\ref{eqn:currspins} and Eqn.~\ref{eqn:convspinpot}, we can derive
\begin{equation}
\begin{pmatrix}
I\\-I_{sL}\\I_{sR}
\end{pmatrix}=-\frac{Ne}{h}\begin{pmatrix}
t & s & s \\
P_r r & \gamma_r & \gamma_t \\
P_t t & \gamma_t & \gamma_r
\end{pmatrix} \begin{pmatrix}
\mu_L-\mu_R \\ \mu_{sL} \\ \mu_{sR}
\end{pmatrix},
\end{equation}
which is Eqn.~2 in the main text, and $\mu_L-\mu_R=-eV$ is induced by a bias voltage $V$. The matrix elements are
\begin{subequations}
	\begin{align}
	t&= t_{\rightarrow\rightarrow} + t_{\rightarrow\leftarrow} + t_{\leftarrow\rightarrow} + t_{\leftarrow\leftarrow},\\
	r&= r_{\rightarrow\rightarrow} + r_{\rightarrow\leftarrow} + r_{\leftarrow\rightarrow} + r_{\leftarrow\leftarrow}=2-t,\\
	\gamma_t &=t_{\rightarrow\rightarrow} - t_{\rightarrow\leftarrow} - t_{\leftarrow\rightarrow} + t_{\leftarrow\leftarrow},\\
	\gamma_r &=r_{\rightarrow\rightarrow} - r_{\rightarrow\leftarrow} - r_{\leftarrow\rightarrow} + r_{\leftarrow\leftarrow}-2,\\
	P_t &= (t_{\rightarrow\rightarrow} - t_{\rightarrow\leftarrow} + t_{\leftarrow\rightarrow} - t_{\leftarrow\leftarrow})/t,\\
	P_r &= (r_{\rightarrow\rightarrow} - r_{\rightarrow\leftarrow} + r_{\leftarrow\rightarrow} - r_{\leftarrow\leftarrow})/r,\\
	s &=t_{\rightarrow\rightarrow} + t_{\rightarrow\leftarrow} - t_{\leftarrow\rightarrow}  -t_{\leftarrow\leftarrow} \\
	&= -r_{\rightarrow\rightarrow} - r_{\rightarrow\leftarrow} + r_{\leftarrow\rightarrow} + r_{\leftarrow\leftarrow}.
	\end{align}
\end{subequations}
For $r=2-t$ and the two expressions of $s$, we have used the condition of charge conservation
\begin{subequations}
	\begin{align}
	t_{\rightarrow\rightarrow} + t_{\rightarrow\leftarrow} + r_{\rightarrow\rightarrow} + r_{\rightarrow\leftarrow}=1,\\
	t_{\leftarrow\rightarrow}+ t_{\leftarrow\leftarrow}+r_{\leftarrow\rightarrow}+r_{\leftarrow\leftarrow}=1.
	\end{align}
\end{subequations}

Among the above matrix elements, $P_t$, $P_r$, and $s$ change sign when the chirality is reversed. For achiral components where there is no spin selective transport, we have $P_t=P_r=s=0$, and the transport matrix is symmetric.

Note that in Eqn.~2 we have defined the vector of currents (thermodynamic response) using $-I_{sL}$ and $I_{sR}$, so that the transport matrix is symmetric in the linear response regime. Following this, as shown in Fig.~1, the chiral component acts as a source (or sink) of spin currents when biased. 

We can rewrite $\gamma_t$ and $\gamma_r$ as transmission-dependent quantities
\begin{subequations}
	\begin{align}
	\gamma_r&=r- (P_r r +s)/\eta_r-2=-t- (P_r r +s)/\eta_r,\\
	\gamma_t&=-t+ (P_t t +s)/\eta_t,
	\end{align}
\end{subequations}
where
\begin{subequations}
	\begin{align}
	\eta_r=\frac{r_{\leftarrow\rightarrow}-r_{\rightarrow\leftarrow}}{r_{\leftarrow\rightarrow}+r_{\rightarrow\leftarrow}},\\
	\eta_t=\frac{t_{\rightarrow\rightarrow}-t_{\leftarrow\leftarrow}}{t_{\rightarrow\rightarrow}+t_{\leftarrow\leftarrow}},
	\end{align}
\end{subequations}
are quantities between $\pm 1$ and change sign under chirality reversal. 

Next, for chiral components where $P_t$, $P_r$, and $s$ are nonzero, the Onsager reciprocity requires the $3 \times 3$ matrix to be symmetric~\cite{onsager1931reciprocal}, which gives
\begin{equation}
	P_t t=P_r r=s,
\end{equation}
and therefore
\begin{subequations}
	\begin{align}
	t_{\rightarrow\leftarrow}&=t_{\leftarrow\rightarrow},\\ r_{\rightarrow\rightarrow}&=r_{\leftarrow\leftarrow},\\ t_{\rightarrow\rightarrow}-t_{\leftarrow\leftarrow}&=r_{\leftarrow\rightarrow}-r_{\rightarrow\leftarrow}.
	\end{align}
\end{subequations}
These expressions show that, for any finite $P_t$ (spin polarization of transmitted electrons), $P_r$ (spin polarization of reflected electrons) must also be nonzero, which then requires the presence of spin-flip reflections~\cite{yang2019spin}. This is again in agreement with the fundamental considerations based on zero charge and spin currents in electrodes at equilibrium.

Further, we obtain
\begin{subequations}
	\begin{align}
	\gamma_r=-(1+2P_t/\eta_r)t,\\
	\gamma_t=-(1-2P_t/\eta_t)t.
	\end{align}
\end{subequations}
which allows us to rewrite the transport matrix equation in terms of $t$
%\begin{widetext}
\begin{equation}\label{eqn:molmatwide}
\begin{pmatrix}
I\\-I_{sL}\\I_{sR}
\end{pmatrix}=-\frac{Ne}{h}\begin{pmatrix}
t & P_t t & P_t t \\
P_t t & -(1+2P_t/\eta_r)t & -(1-2P_t/\eta_t)t \\
P_t t & -(1-2P_t/\eta_t)t & -(1+2P_t/\eta_r)t
\end{pmatrix} \begin{pmatrix}
\mu_L-\mu_R \\ \mu_{sL} \\ \mu_{sR}
\end{pmatrix}.
\end{equation}
%\end{widetext}
Note that $P_t$, $\eta_t$, and $\eta_r$ all change sign when the chirality is reversed, and $t$, $P_t/\eta_t$, and $P_t/\eta_r$ are always positive.

In later discussions we connect the right-hand side of the nonmagnetic component to an electrode (see Fig.~2(a)) where the spin accumulation $\mu_{sR}$ is zero and the spin current $I_{sR}$ is irrelevant. This reduces the matrix equation to
\begin{equation}\label{eqn:molmatminsup}
\begin{pmatrix}
I\\-I_{sL}
\end{pmatrix}=-\frac{Ne}{h}\begin{pmatrix}
t & P_t t \\
P_t t & -(1+2P_t/\eta_r)t
\end{pmatrix} \begin{pmatrix}
\mu_L-\mu_R \\ \mu_{sL} 
\end{pmatrix},
\end{equation}
which is equivalent to Eqn.~3 in the main text.

\section{E. Spin and charge transport at an achiral ferromagnetic tunnel junction/interface}
In a FM, time-reversal symmetry is broken and the Kramers degeneracy is lifted. The FMTJ provides a spin polarization to any current that flows through, and reciprocally, it generates a charge voltage upon a spin accumulation at its interface. In terms of spin and charge transport, the FMTJ allows different conductances for electrons with opposite spins, and the difference depends on the spin polarization $P_{FM}$. Inside the FM spin relaxation is strong and the spin accumulation is considered zero. For our discussions, the FM is always connected to the left of the node, we thus only consider the $R$ interface of the FM with a spin accumulation $\mu_{sR}$. Following the discussion provided in Ref.~\onlinecite{jansen2018nonlinear}, the spin-specific currents on the $R$-side are therefore  
\begin{subequations}
	\begin{align}
	I_{\rightarrow}=-\frac{1}{e}G_{\rightarrow}[\mu_L-(\mu_R + \mu_{sR})],\\
	I_{\leftarrow}=-\frac{1}{e}G_{\leftarrow}[\mu_L-(\mu_R - \mu_{sR})],
	\end{align}
\end{subequations}
where $G_{\rightarrow(\leftarrow)}$ is the spin-specific tunnel conductance. We define the total conductance $G_{FM}=G_{\rightarrow}+G_{\leftarrow}$ and FM spin-polarization $P_{FM}=(G_{\rightarrow}-G_{\leftarrow})/(G_{\rightarrow}+G_{\leftarrow})$. We also introduce a transmission coefficient $T$ ($0\leqslant T \leqslant 2$) to distinguish $G_{FM}$ from ideal transmission. The charge and spin currents on the $R$ interface of the FM can thus be written as
\begin{equation}
\begin{pmatrix}
I\\I_{sR}
\end{pmatrix}=-\frac{N'e}{h}\begin{pmatrix}
T & -P_{FM} T \\
P_{FM} T & -T
\end{pmatrix} \begin{pmatrix}
\mu_L -\mu_R \\ \mu_{sR}
\end{pmatrix},
\end{equation}
where $N'$ is the number of spin-degenerate channels in the FMTJ.

This matrix gives the spin and charge response of the FMTJ, and unlike for a chiral molecule, this matrix fulfills the Onsager reciprocity by being antisymmetric (opposite off-diagonal terms), because upon magnetic field and magnetization reversal, the FM spin polarization $P_{FM}$ changes sign~\cite{onsager1931reciprocal}. 

\section{F. Spin and charge transport in a generic 2T circuit}
We consider a generic 2T circuit like the one shown in Fig.~2(a). Our goal here is to derive the quantities $I$, $\mu$, and $\mu_s$ as a function of 2T bias $\mu_L-\mu_R$. We rewrite the transport matrices into conductance matrices, and obtain for the left and right side of the node
\begin{subequations}
	\begin{align}
	\begin{pmatrix}
	I \\ I_s
	\end{pmatrix}
	=-\frac{1}{e}\begin{pmatrix}
	G_1 & G_2 \\ G_3 & G_4
	\end{pmatrix} \begin{pmatrix}
	\mu_L-\mu \\ \mu_s
	\end{pmatrix}, \\
	\begin{pmatrix}
	I \\ -I_s
	\end{pmatrix}
	=-\frac{1}{e}\begin{pmatrix}
	g_1 & g_2 \\ g_3 & g_4
	\end{pmatrix} \begin{pmatrix}
	\mu - \mu_R \\ \mu_s
	\end{pmatrix}.
	\end{align}
\end{subequations}

At steady state the continuity condition requires the currents $I$ and $I_s$ are equal for both equations. This condition allows us to derive
\begin{subequations}
	\begin{align}
	\begin{split}
	\mu&=\mu_R +\frac{G_3 (G_2-g_2)-G_1 (G_4+g_4)}{f} ~\left(\mu_L - \mu_R\right)\\
	&=\mu_L -\frac{g_3 (g_2-G_2)-g_1 (G_4+g_4)}{f} ~\left(\mu_L - \mu_R\right),
	\end{split}\\
	\mu_s&=\frac{(G_3/G_1 + g_3/g_1) G_1 g_1}{f} ~\left(\mu_L - \mu_R \right),\\
	I&=\frac{G_1 g_2 g_3 + g_1 G_2 G_3 - G_1 g_1 (G_4 + g_4)}{f} ~\left(\mu_L - \mu_R \right),
	\end{align}
\end{subequations} 
where the coefficient $f$ is
\begin{equation}
f=(G_3 - g_3)(G_2 - g_2) - (G_1 + g_1)(G_4 + g_4).
\end{equation}

Notably, $f$ depends on $(G_3 - g_3)(G_2 - g_2)$, and for our linear-regime example using a FM with $G_2=-G_3$ and a chiral component with $g_2=g_3$, $f$ only depends on $g_3^2-G_3^2$, which does not change when either the FM magnetization or the chirality is reversed. Moreover, the charge current $I$ only depends on $g_3^2$ and $G_3^2$, and therefore remains unchanged under magnetization or chirality reversal. In contrast, the spin accumulation $\mu_s$ depends on $G_3$ and $g_3$, and the charge distribution indicated by $\mu$ depends on $G_3 g_2$ or $g_3 G_2$. Both $\mu_s$ and $\mu$ therefore indeed change upon magnetization or chirality reversal, and can be detected using three-terminal or four-terminal measurements.

If the two circuit components would both be FM (antisymmetric $G$ and $g$ matrices) or both be chiral (symmetric $G$ and $g$ matrices), then the coefficient $f$ contains the cross products $G_3 g_2$ and $G_2 g_3$, and therefore gives rise to a 2T current that does change upon magnetization or chirality reversal.  

\section{G. Reciprocity breaking by energy-dependent tunneling}
Here we derive the energy dependence of the charge and spin tunneling through the FM interface following the discussion by Jansen \textit{et al.}~\cite{jansen2018nonlinear}, and derive the subsequent nonlinear coupled charge and spin transport equations.

\begin{figure}[hbt]
	\includegraphics[width=0.55\linewidth]{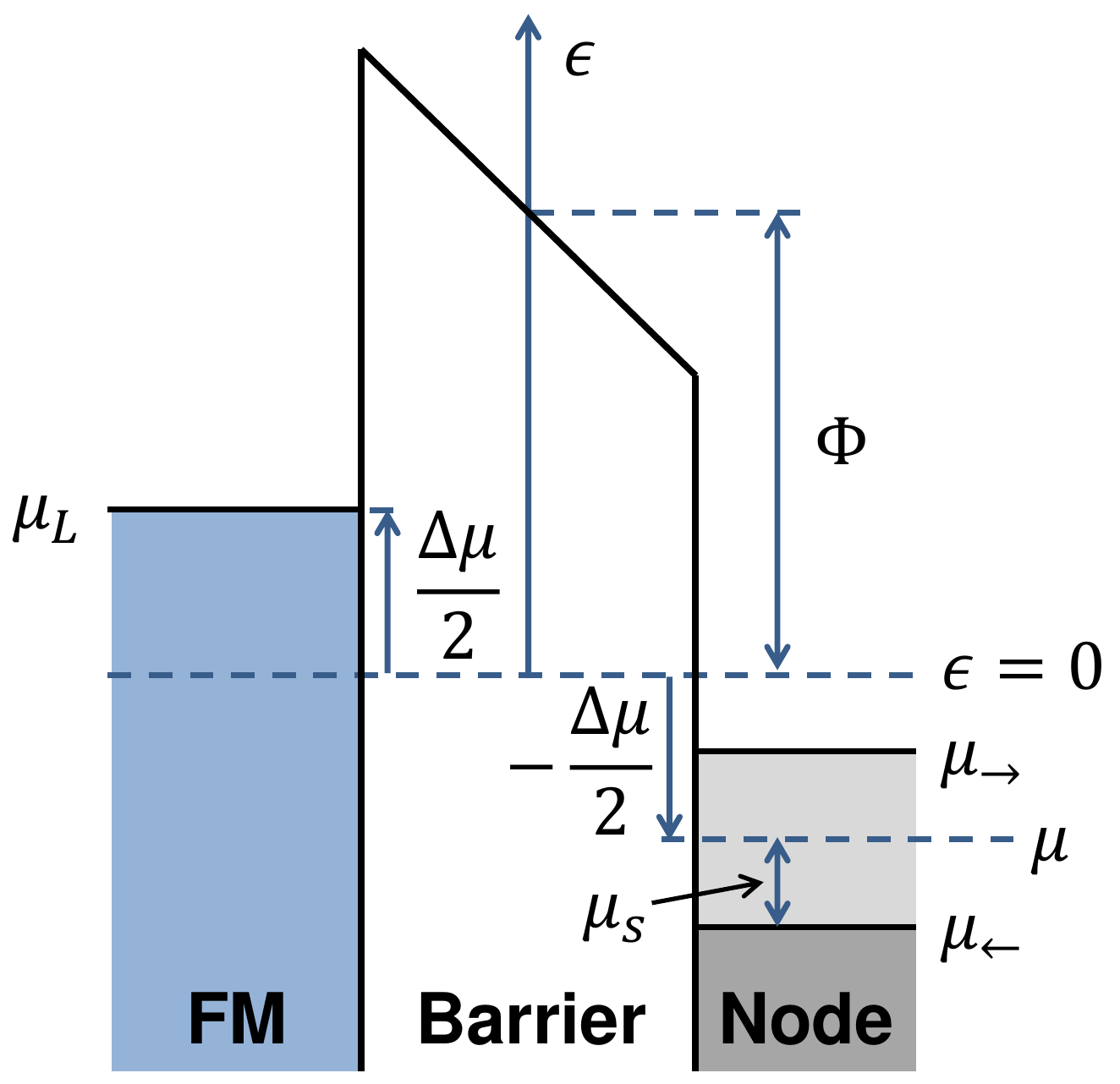}
	\caption{\label{fig:barrierfm}Energy diagram of the tunnel barrier at the ferromagnet interface. Details see text.}
\end{figure}

We consider a biased symmetric rectangular tunnel barrier with height $\Phi$ and width $w$ between the FM and the node, as shown in Fig.~\ref{fig:barrierfm}. The reference energy $\epsilon=0$ is chosen as the average of $\mu_L$ and $\mu$, so that the average barrier height does not change with bias. The bias voltage $V=-\Delta \mu/e$ is applied, so that $\mu_L=\Delta\mu/2$, $\mu=-\Delta\mu/2$, $\mu_{\rightarrow}=-\Delta\mu/2+\mu_s$, and $\mu_{\leftarrow}=-\Delta\mu/2-\mu_s$. For simplicity, we assume the barrier height (evaluated at the center of the tunnel barrier) is constant, the tunneling is one dimensional, and the electrostatic potential drop across the tunnel barrier is exactly $-eV$.

The energy-dependent electron transmission function is
\begin{equation}\label{eqns:transbarrier}
T(\epsilon)=2 \exp \left( -\beta~(\Phi-\epsilon)^{1/2} \right),
\end{equation}
where $\beta=2w \sqrt{2m/\hbar^2}$, with $m$ being the (effective) electron mass and $\hbar$ the reduced Planck's constant. 

The total transmitted current is an energy-integral of two energy-dependent functions: (1). the transmission function $T(\epsilon)$, and (2) the voltage-dependent Fermi-Dirac distribution functions in both electrodes. The two parts dominate at different bias regimes with respect to the thermal activation energy $k_B T$, where $k_B$ is the Boltzmann constant and $T$ is temperature. For tunneling, typically we have $|eV| \gg k_B T$, and therefore we assume zero temperature for convenience. The Fermi-Dirac distribution functions then become Heaviside step functions with the step at $\mu_L$ for the left electrode and at $\mu_{\rightarrow}$ and $\mu_{\leftarrow}$ for the two spin species in the node.

Assuming $P_{FM}$ does not depend on energy, the energy-integrated tunnel current for each spin species through the FMTJ is
\begin{subequations}
	\begin{align}
	I_{\rightarrow}(\Delta\mu, \mu_s)=-\frac{N'e}{h}~ \frac{1+P_{FM}}{2}~\int_{\mu_{\rightarrow}}^{\mu_L}T(\epsilon)~d\epsilon,\\
	I_{\leftarrow}(\Delta\mu, \mu_s)=-\frac{N'e}{h}~\frac{1-P_{FM}}{2}~ \int_{\mu_{\leftarrow}}^{\mu_L}T(\epsilon)~d\epsilon.
	\end{align}
\end{subequations}
The range of the integrals can be rewritten as
\begin{subequations}
	\begin{align}
\text{for}~I_{\rightarrow}:&~~\int_{\mu_{\rightarrow}}^{\mu_L} d\epsilon= \int_{\mu}^{\mu_L} d\epsilon - \int_{\mu}^{\mu_{\rightarrow}} d\epsilon, \\
\text{for}~I_{\leftarrow}:&~~\int_{\mu_{\leftarrow}}^{\mu_L} d\epsilon= \int_{\mu}^{\mu_L} d\epsilon + \int_{\mu_{\leftarrow}}^{\mu} d\epsilon.
	\end{align}
\end{subequations}
We denote the two integrals on the right-hand side of the expressions $I_{\rightarrow(\leftarrow)}^{(1)}$ and $I_{\rightarrow(\leftarrow)}^{(2)}$ respectively, so that we have $I_{\rightarrow}=I_{\rightarrow}^{(1)}+I_{\rightarrow}^{(2)}$ and $I_{\leftarrow}=I_{\leftarrow}^{(1)}+I_{\leftarrow}^{(2)}$. The first integral $I_{\rightarrow(\leftarrow)}^{(1)}$ depends on the electrochemical potential difference across the barrier, $\Delta \mu$, and thus describes the bias-induced transport. In contrast, the second integral $I_{\rightarrow(\leftarrow)}^{(2)}$ depends on $\mu$ and $\mu_s$ in the node, and it vanishes when $\mu_s=0$, and therefore it describes the spin-accumulation-induced correction to the first integral. We will treat the two integrals separately. 

The bias-induced charge and spin currents are
\begin{subequations}\label{eqns:Ibias}
	\begin{equation}
	\begin{split}
	I^{(1)}&=I_{\rightarrow}^{(1)}+I_{\leftarrow}^{(1)}\\&=-\frac{N'e}{h}\int_{\mu}^{\mu_L}T(\epsilon)~d\epsilon\\&=-\frac{N'e}{h}~\left(\overline{T}|_{\mu}^{\mu_L}\right)~\Delta\mu,
	\end{split}
	\end{equation}
	\begin{equation}
	\begin{split}
	I_s^{(1)}&=I_{\rightarrow}^{(1)}-I_{\leftarrow}^{(1)}\\&=-\frac{N'e}{h}~P_{FM}~\int_{\mu}^{\mu_L}T(\epsilon)~d\epsilon\\&=-\frac{N'e}{h}~\left(P_{FM}~\overline{T}|_{\mu}^{\mu_L}\right)~\Delta\mu,
	\end{split}
	\end{equation}
\end{subequations}
where the averaged transmission within the bias window is defined as 
\begin{equation}
\overline{T}|_{\mu}^{\mu_L}=\frac{1}{\Delta\mu}~\int_{\mu}^{\mu_L}~T(\epsilon)~d\epsilon.
\end{equation}
Note that when $\Delta \mu$ changes sign, the integral changes sign too, and therefore $\overline{T}|_{\mu}^{\mu_L}$ is an even function of $\Delta \mu$.

For the spin-accumulation-induced currents, we have
\begin{subequations}\label{eqns:Ispinaccu}
	\begin{equation}
	\begin{split}
	&I^{(2)}=I_{\rightarrow}^{(2)}+I_{\leftarrow}^{(2)}\\&=-\frac{N'e}{h} \left[ -\frac{1+P_{FM}}{2}\int_{\mu}^{\mu_{\rightarrow}}T(\epsilon)d\epsilon +\frac{1-P_{FM}}{2}\int_{\mu_{\leftarrow}}^{\mu}T(\epsilon)d\epsilon \right]\\
	&\approx -\frac{N'e}{h} \left(-P_{FM}T|_{\epsilon=\mu}\right)\cdot\mu_s,
	\end{split}
	\end{equation}
	\begin{equation}
	\begin{split}
	&I_s^{(2)}=I_{\rightarrow}^{(2)}-I_{\leftarrow}^{(2)}\\&=-\frac{N'e}{h} \left[ -\frac{1+P_{FM}}{2}\int_{\mu}^{\mu_{\rightarrow}}T(\epsilon)d\epsilon -\frac{1-P_{FM}}{2}\int_{\mu_{\leftarrow}}^{\mu}T(\epsilon)d\epsilon \right]\\
	&\approx -\frac{N'e}{h} \left(-T|_{\epsilon=\mu}\right)\cdot\mu_s,
	\end{split}
	\end{equation}
\end{subequations}
where the approximation is taken under the assumption that $\mu_s \ll \Delta \mu$, so that $T(\epsilon)$ is approximately a constant within the energy range from $\mu_{\leftarrow}$ to $\mu_{\rightarrow}$, and thus they are evaluated at $\epsilon=\mu =-\Delta\mu/2$. The energy dependence of $I^{(2)}$ and $I_s^{(2)}$ follows that of $T(\epsilon)$ (Eqn.~\ref{eqns:transbarrier}).

We can now rewrite Eqn.~\ref{eqns:Ibias} and Eqn.~\ref{eqns:Ispinaccu} into matrix form
\begin{equation}
\begin{pmatrix}
I\\ \hspace{-0.3cm} \\I_s
\end{pmatrix}=-\frac{N'e}{h}\begin{pmatrix}
\overline{T}|_{\mu}^{\mu_L} & -P_{FM}T|_{\epsilon=\mu} \\ \hspace{-0.3cm} \\
P_{FM}\overline{T}|_{\mu}^{\mu_L} & -T|_{\epsilon=\mu}
\end{pmatrix} \begin{pmatrix}
\mu_L - \mu \\ \hspace{-0.3cm} \\ \mu_s
\end{pmatrix},
\end{equation}
which is Eqn.~6 in the main text. The different off-diagonal terms demonstrate the breaking of Onsager reciprocity in the nonlinear regime, since one of them depends on the integral of the transmission function, while the other depends on the transmission function itself. In the linear response regime ($\mu_L \approx \mu$), the two terms differ only by sign.

\begin{figure}[hbt]
	\includegraphics[width=0.55\linewidth]{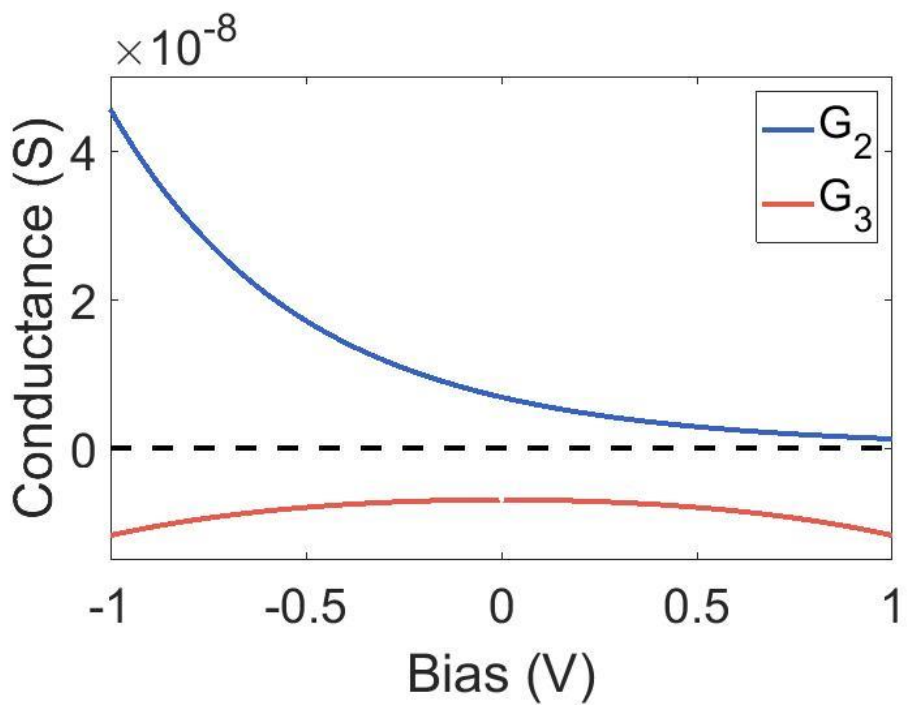}
	\caption{\label{fig:parafm}Bias dependence of the off-diagonal elements of the conductance matrix describing tunneling through the FMTJ.}
\end{figure}

We write
\begin{equation}
\begin{pmatrix}
I \\ I_s
\end{pmatrix}
=-\frac{1}{e}\begin{pmatrix}
G_1 & G_2 \\ G_3 & G_4
\end{pmatrix} \begin{pmatrix}
\mu_L-\mu \\ \mu_s
\end{pmatrix},
\end{equation}
so that the matrix elements represent conductance, where $G_2= (N'e^2/h) \cdot (-P_{FM}T|_{\epsilon=\mu})$ and $G_3=(N'e^2/h) \cdot (P_{FM}\overline{T}|_{\mu}^{\mu_L})$. In Fig.~\ref{fig:parafm}, we plot $G_2$ and $G_3$ as a function of bias across the FMTJ ($(\mu_L-\mu)/e$) for $P_{FM}=-0.5$ and $N'=1000$, values that are used to calculate the $I$-$V$ curves in Fig.~2(d) (for other parameters see later in SI.I). 

Note that $G_2=-G_3$ at zero bias, as required by Onsager reciprocity in the linear response regime. The breaking of reciprocity in the nonlinear regime is exhibited by the unequal absolute values of $G_2$ and $G_3$ away from zero bias. Here $G_2$ is increases monotonically with decreasing bias, while $G_3$ is symmetric in bias and is minimum at zero bias.

\section{H. Reciprocity breaking by thermally activated resonant transmission through molecular orbitals}
\begin{figure}[hbt]
	\includegraphics[width=0.55\linewidth]{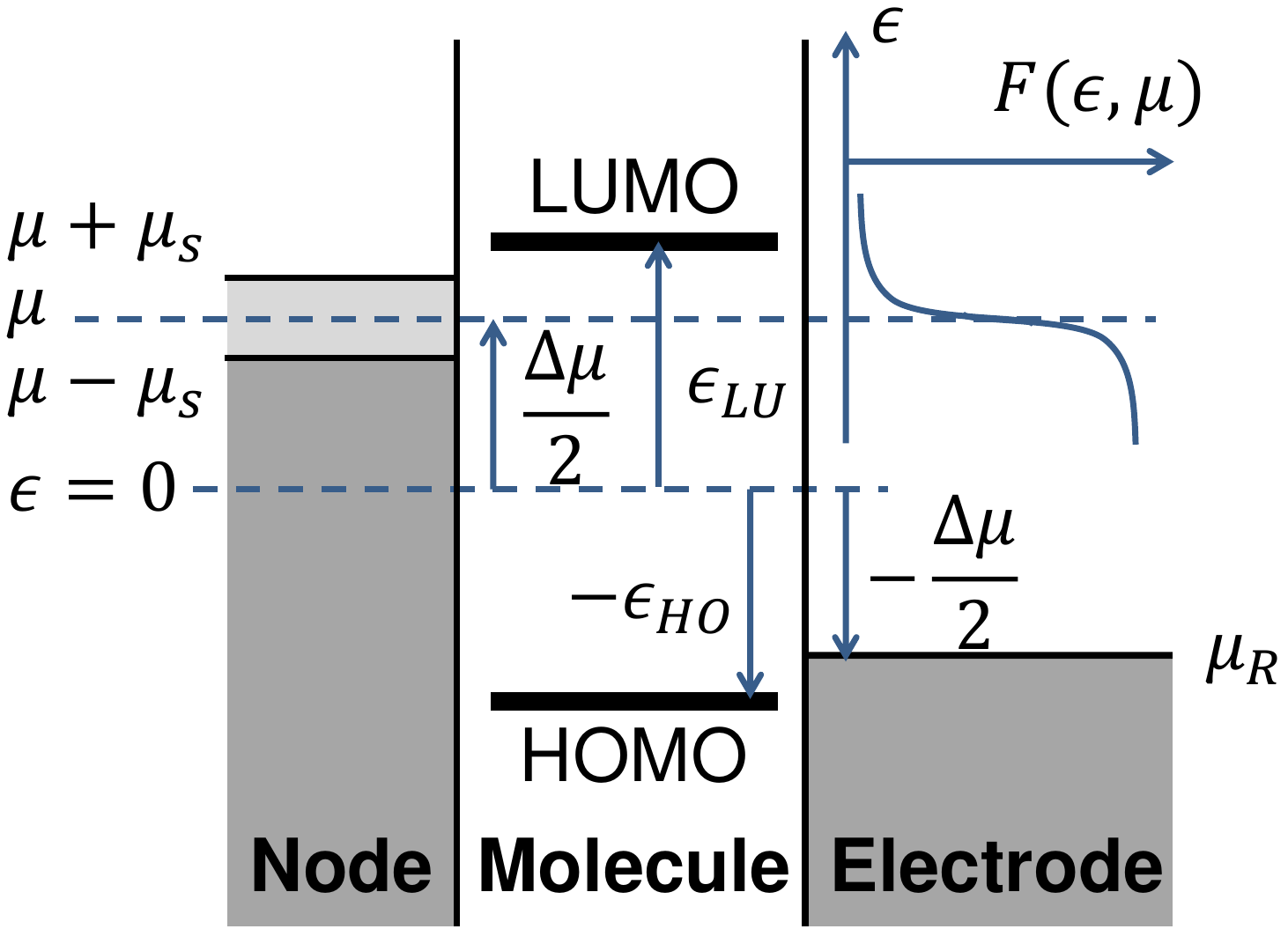}
	\caption{\label{fig:moltrans}Energy diagram of the electron transmission through molecular levels.}
\end{figure}

We discuss here the effect of the Fermi-Dirac distribution of electrons using resonant transmission through discrete molecular orbitals (levels), specifically the lowest-unoccupied molecular orbital (LUMO) and the highest-occupied molecular orbital (HOMO). This is illustrated in Fig.~\ref{fig:moltrans}. We assume here the LUMO and HOMO levels are fixed at energies $\epsilon_{LU}$ and $-\epsilon_{HO}$ respectively ($\epsilon_{LU}, \epsilon_{HO}>0$), with $\epsilon=0$ defined as the average of the two (charge) electrochemical potentials on both sides of the molecule. The nonlinearity is now assumed to be fully due to the Fermi-Dirac distribution $F(\epsilon,\mu)$, which is different for the node and the electrode, and is also different for the two spin species in the node when a spin accumulation is considered.

Consider the geometry in Fig.~\ref{fig:moltrans}, at a given bias $\Delta \mu$ and a spin accumulation $\mu_s$ in the node, the spin-specific Fermi-Dirac functions for the two spin species in the node are
\begin{subequations}
	\begin{align}
	F_{L\rightarrow}(\epsilon,\Delta\mu,\mu_s)=\frac{1}{\exp \left( \frac{\epsilon-\Delta\mu/2-\mu_s}{k_B T} \right)+1},\\
	F_{L\leftarrow}(\epsilon,\Delta\mu,\mu_s)=\frac{1}{\exp \left( \frac{\epsilon-\Delta\mu/2+\mu_s}{k_B T} \right)+1},
	\end{align}
\end{subequations}
where we used subscript $L$ to denote the node because its located on the left side of the molecule.

The Fermi-Dirac function in the node (neglecting the spin accumulation) and in the right electrode are
\begin{subequations}
	\begin{align}
	F_L(\epsilon, \Delta\mu)=\frac{1}{\exp \left( \frac{\epsilon-\Delta\mu/2}{k_B T} \right)+1},\\
	F_R(\epsilon, \Delta\mu)=\frac{1}{\exp \left( \frac{\epsilon+\Delta\mu/2}{k_B T} \right)+1}.
	\end{align}
\end{subequations}

We model the resonant transmission through the molecular orbitals as energy-dependent transmission probability $t(\epsilon)$ that is only nonzero at the LUMO and the HOMO levels, which it is described by
\begin{equation}
t(\epsilon)=t_{LU} \delta(\epsilon-\epsilon_{LU})+t_{HO} \delta(\epsilon+\epsilon_{HO}),
\end{equation}
where $t_{LU}$ and $t_{HO}$ are the transmission probabilities through the LUMO and the HOMO, respectively, and $\delta(\epsilon)$ is a Dirac delta function.

An exact calculation should separately address the two spin orientations separately, but for our assumption of $\mu_s \ll \Delta \mu$, we will directly start from the linear-regime equation of coupled charge and spin transport (Eqn.~\ref{eqn:molmatminsup})
\begin{equation}
\begin{pmatrix}
I\\-I_{sL}
\end{pmatrix}=-\frac{Ne}{h}\begin{pmatrix}
t & P_t t \\
P_t t & -(1+2P_t/\eta_r)t
\end{pmatrix} \begin{pmatrix}
\mu-\mu_R \\ \mu_{s}
\end{pmatrix}.
\end{equation}
This gives the charge current $I$
\begin{equation}
I=-\frac{Ne}{h} \left[t(\mu_L-\mu_R) + P_t t (\mu_{L\rightarrow}-\mu_{L\leftarrow})/2 \right],
\end{equation}
where we used $\mu_{L\rightarrow}=\mu+\mu_s$ and $\mu_{L\leftarrow}=\mu-\mu_s$.

At finite temperature, the two terms in the square bracket on the right-hand side of the equation should each be replaced by an energy-integral of the electron transmission function modified by the voltage- and temperature-dependent Fermi-Dirac distributions in both electrodes. We denote the two terms $I^{(1)}$ and $I^{(2)}$, and the first term is
\begin{equation}
\begin{split}
I^{(1)}&=\int_{-\infty}^{\infty} t(\epsilon) \left[F_L(\epsilon,\Delta\mu)-F_R(\epsilon,\Delta\mu) \right] d\epsilon\\
&=t_{LU} \left[F_L(\epsilon_{LU},\Delta\mu)-F_R(\epsilon_{LU},\Delta\mu) \right]\\
&~~~+t_{HO} \left[F_L(-\epsilon_{HO},\Delta\mu)-F_R(-\epsilon_{HO},\Delta\mu) \right].	
\end{split}
\end{equation} 
We denote
\begin{equation}
\mathcal{F}_{\epsilon}(\Delta \mu)=\frac{1}{\Delta\mu} \left[F_L(\epsilon,\Delta\mu)-F_R(\epsilon,\Delta\mu)\right],
\end{equation}
to highlight the transmission at energy $\epsilon$ and under bias $\Delta\mu$. In this manner, we can write 
\begin{equation}
I^{(1)}=\left[t_{LU}~\mathcal{F}_{\epsilon_{LU}}(\Delta \mu)+t_{HO}~\mathcal{F}_{-\epsilon_{HO}}(\Delta \mu)\right]~\Delta\mu.
\end{equation}

Similarly, at finite temperature, the second term $I^{(2)}$ becomes
\begin{equation}
\begin{split}
&I^{(2)}\\
=&\frac{P_t}{2} (t_{LU} \left[F_{L\rightarrow}(\epsilon_{LU},\Delta\mu, \mu_s)-F_{L\leftarrow}(\epsilon_{LU},\Delta\mu,\mu_s) \right]\\
&~+t_{HO} \left[F_{L\rightarrow}(-\epsilon_{HO},\Delta\mu, \mu_s)-F_{L\leftarrow}(-\epsilon_{HO},\Delta\mu,\mu_s) \right]).
\end{split}
\end{equation}

We define
\begin{equation}
\begin{split}
&\mathcal{F}'_{L,\epsilon}(\Delta\mu)\\
=&\frac{1}{2\mu_s}~\left[F_{L\rightarrow}(\epsilon,\Delta\mu, \mu_s)-F_{L\leftarrow}(\epsilon,\Delta\mu,\mu_s)\right]\\
\approx &
\frac{\partial [F_L(\epsilon,\Delta\mu)]}{\partial\Delta\mu},\\
\end{split}
\end{equation}
where the approximation is taken under the assumption of $\mu_s \ll \Delta \mu$. 

With this, we have
\begin{equation}
I^{(2)}=P_t \left[t_{LU}~\mathcal{F}'_{L,\epsilon_{LU}}(\Delta\mu)+t_{HO}~\mathcal{F}'_{L,-\epsilon_{HO}}(\Delta\mu)\right]\mu_{s}.
\end{equation}

Similarly, the expression of $-I_{sL}$ at finite temperature can also be derived from the linear regime expression. The nonlinear coupled spin and charge transport equation, due to the thermally activated resonant transmission via the LUMO and the HOMO levels, is therefore
%\begin{widetext} 
\begin{equation}
\begin{pmatrix}
I\\-I_{sL}
\end{pmatrix}=-\frac{Ne}{h}\begin{pmatrix}
t_{LU}~\mathcal{F}_{\epsilon_{LU}}(\Delta \mu)+t_{HO}~\mathcal{F}_{- \epsilon_{HO}}(\Delta \mu) & P_t [t_{LU}~\mathcal{F}'_{L,\epsilon_{LU}}(\Delta \mu)+t_{HO}~\mathcal{F}'_{L,-\epsilon_{HO}}(\Delta \mu)] \\
P_t [t_{LU}~\mathcal{F}_{\epsilon_{LU}}(\Delta \mu)+t_{HO}~\mathcal{F}_{-\epsilon_{HO}}(\Delta \mu)] & -(1+2P_t/\eta_r)~[t_{LU}~\mathcal{F}'_{L,\epsilon_{LU}}(\Delta \mu)+t_{HO}~\mathcal{F}'_{L,-\epsilon_{HO}}(\Delta \mu)]
\end{pmatrix} \begin{pmatrix}
\mu-\mu_R \\ \mu_{s}
\end{pmatrix},
\end{equation}
%\end{widetext}
where we have assumed $P_t$, $\eta_r$ do not depend on energy.

The breaking of reciprocity arises from the different forms of the off-diagonal terms of the transport matrix. While the bottom-left term scales with the electron occupation function, the top-right terms scales with its derivative. In the linear response regime, the two terms are equal and Onsager reciprocity is present.

If required, this nonlinear form can be easily extended to the $3\times 3$ matrix for a generalized situation considering also a spin accumulation $\mu_{sR}$ on the right-hand side of the molecule. Then the third column will be modified similarly to the second column using the derivative of the electron occupation function in the right electrode $F_R(\epsilon,\Delta\mu)$.

\begin{figure}[hbt]
	\includegraphics[width=0.55\linewidth]{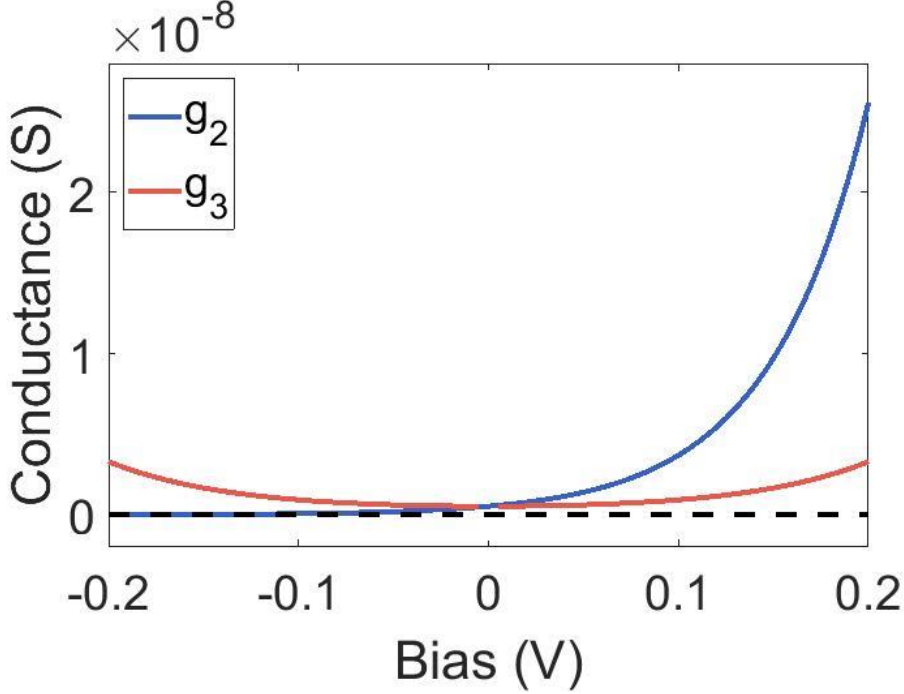}
	\caption{\label{fig:paramol}Bias dependence of the off-diagonal elements of the conductance matrix describing thermally activated conduction through the LUMO.}
\end{figure}

Similar to the FMTJ, we can write the above equation into 
\begin{equation}
\begin{pmatrix}
I \\ -I_s
\end{pmatrix}
=-\frac{1}{e}\begin{pmatrix}
g_1 & g_2 \\ g_3 & g_4
\end{pmatrix} \begin{pmatrix}
\mu - \mu_R \\ \mu_s
\end{pmatrix},
\end{equation}
where the matrix elements represent conductances. We plot the off-diagonal terms as a function of bias ($(\mu -\mu_R)/e$) for transmission only through the LUMO level ($t_{LU}=1$, $t_{HO}=0$, $\epsilon_{LU}=0.6$~eV) for $P_t=0.85$ and $N=1000$, as shown in Fig.~\ref{fig:paramol}. 

Note that $g_2=g_3$ at zero bias, as required by Onsager reciprocity in the linear response regime. The reciprocity is broken as bias shifts away from zero, and the two conductances are no longer equal. Here $g_2$ increases monotonically with bias, while $g_3$ is bias-symmetric and reaches minimum at zero bias.

\section{I. Parameters for example $I$-$V$ curves}
(1) For calculating the $I$-$V$ curves in Fig.~2(d) we have used the following parameters: Energy dependent FM tunnel transmission $T(\epsilon)$ for a tunnel barrier with height $\Phi=2.1$~eV, and width $w=1.0$~nm. The number of channels are set as $N'=1000$ and $N=5000$, the values are chosen taking into account the device length scale. For example, if using Ni as FM, when assuming each atom provides one channel, $N'=1000$ corresponds to roughly an area of $100$~nm$^2$. The transmission through the chiral component is set linear and uses the same transmission value as the FM tunnel barrier at zero bias. The polarization values are $P_{FM}=\pm0.5$, $P_t=0.85$, and the quantity $\eta_r$ is set as $\eta_r=0.9$.
The same parameters are used to obtain the conductances in Fig.~\ref{fig:parafm}.
 
(2) For calculating the $I$-$V$ curves in Fig.~3(b-c) we have used the following parameters: Temperature at $300$~K, the LUMO level at $\epsilon_{LU}=0.6$~eV, and the HOMO level at $-\epsilon_{HO}=-0.6$~eV. For Fig.~3(b), the LUMO transmission $t_{LU}=1$ and the HOMO transmission $t_{HO}=0$, while for Fig.~3(c) the two values are interchanged. The number of channels are set as $N'=1000$ and $N=200$. The transmission through the FM tunnel barrier is set linear and uses the same transmission value as the FM tunnel barrier in case (1) at zero bias. The polarization values are $P_{FM}=\pm0.5$, $P_t=0.85$, and the quantity $\eta_r$ is set as $\eta_r=0.9$. The same parameters are used to obtain the conductances in Fig.~\ref{fig:paramol}.

(3) For the chiral spin valve in Fig.~4(f), we use the same tunnel transmission function and the same barrier parameters as in case (1) for both chiral components (both nonlinear), and do not consider the thermally activated molecular level transmission. The number of channels are set as $N'=1000$ and $N=20000$. The first chiral component has polarization $\pm0.5$, and the second one $0.85$, both have $\eta_r=0.9$.

(d) For the example in Fig.~4(g), the tunnel barrier uses the same parameters as the FMTJ in case (1), but has zero spin polarization. The chiral component transmission is set linear and uses the same set of parameters as in case (1). The dashed curve is obtained by forcing $\mu_s = 0$ for all bias values. 

\section{J. Sign of the nonlinear 2T MR}
We summarize here, separately for the two nonlinear mechanisms, how the sign of the MR depends on the chirality, the charge carrier type, and the bias direction. First of all, we define the MR for the device in Fig.~2(a) following 
\begin{equation}
MR=\frac{I_+-I_-}{I_++I_-} \times 100\%,
\end{equation}
where $I_+$ represents the charge current measured with the FM polarization $P_{FM}>0$, which corresponds to a spin-right polarization of the injected electrons, and $I_-$ corresponds to the current measured when the FM generates a spin-left polarization.

For the bias direction, we describe it in terms of electron movement direction (opposite to charge current) with respect to the circuit components. For example, in Fig.~2(a), the positive bias voltage corresponds to an electron movement from the chiral component (CC) to the FMTJ.

For the (sign of) chirality, we label it as \textsc{d} and \textsc{l}. Here we (arbitrarily) assume that a \textsc{d}-chiral component favors the transmission of electrons with spins parallel to the electron momentum, while a \textsc{l}-chiral component favors the anti-parallel. The exact correspondence between the chirality and the favored spin orientation depends on the microscopic details of the spin orbit interaction. 

According to these definitions, the sign of the MR is summarized in the following tables.

\begin{table}[!h]
	\caption{Summary of the sign of MR for nonlinear tunneling through FMTJ}
\begin{tabular}{c|c|c||c}
	\hline
	chirality & electron direction & carrier type & MR\\
	\hline
	\textsc{d}      &  CC to FMTJ &   electron  & +  \\
	\textsc{l}      &  CC to FMTJ &   electron  & -\\
	\textsc{d}      &  FMTJ to CC &   electron  & -\\
	\textsc{l}      &  FMTJ to CC &   electron  & +\\
	\hline
\end{tabular}
\end{table}

\begin{table}[!h]
	\caption{Summary of the sign of MR for nonlinear transmission through molecular orbitals}
	\begin{tabular}{c|c|c||c}
		\hline
		chirality & electron direction & carrier type & MR\\
		\hline
		\textsc{d}      &  CC to FMTJ &   electron  & +  \\
		\textsc{l}      &  CC to FMTJ &   electron  & -\\
		\textsc{d}      &  FMTJ to CC &   electron  & -\\
		\textsc{l}      &  FMTJ to CC &   electron  & +\\
		\textsc{d}      &  CC to FMTJ &   hole  & -\\
		\textsc{l}      &  CC to FMTJ &   hole  & +\\
		\textsc{d}      &  FMTJ to CC &   hole  & +\\
		\textsc{l}      &  FMTJ to CC &   hole  & -\\
		\hline
	\end{tabular}
\end{table}